\documentclass[epjCONF]{svjour}
\usepackage{graphicx}
\usepackage[varg]{txfonts} 
\usepackage[latin1]{inputenc}
\usepackage{axodraw}
\session-title{MESON2012 - 12th International Workshop on Meson Production, Properties and Interaction}
\begin{document}
\title{The hadronic light-by-light contribution to the muon anomalous magnetic moment and renormalization group for EFT}
\author{Johan Bijnens\thanks{speaker, \email{bijnens@thep.lu.se}} \and 
Mehran Zahiri Abyaneh}
\institute{Dept. of Astronomy and Theoretical Physics, Lund University,
S\"olvegatan 14A, 22362 Lund, Sweden}
\abstract{
We give a short overview of the theory
of the muon anomalous magnetic moment
with emphasis on the hadronic light-by-light and the pion loop contribution.
We explain the difference between
the hidden local symmetry and full VMD pion loop and
discuss leading logarithms in the anomalous sector of
2-flavour chiral perturbation theory.
} 

\thispagestyle{empty}
\begin{flushright}
\large
LU TP 12-32\\
August 2012
\end{flushright}

\vfill

\begin{center}
{\huge\bf
The hadronic light-by-light contribution to the muon anomalous magnetic moment and renormalization group for EFT$^*$}\\[1.5cm]
{\large\bf Johan Bijnens and 
Mehran Zahiri Abyaneh}\\[1cm]
{\large Dept. of Astronomy and Theoretical Physics, Lund University,
S\"olvegatan 14A, 22362 Lund, Sweden}

\vfill

{\large\bf Abstract}
\end{center}

We give a short overview of the theory
of the muon anomalous magnetic moment
with emphasis on the hadronic light-by-light and the pion loop contribution.
We explain the difference between
the hidden local symmetry and full VMD pion loop and
discuss leading logarithms in the anomalous sector of
2-flavour chiral perturbation theory.

\vfill
\noindent\rule{8cm}{0.5pt}\\
$^*$ Invited talk presented by JB at MESON2012 - 12th International Workshop on Meson Production, Properties and Interaction, Krakow, Poland, 31 May -- 5 June 2012
\setcounter{page}{0}
\def\epjrunnhead{\markboth{submitted to EPJ Web of Conferences}{MESON2012 - 12th International Workshop on Meson Production, Properties and Interaction}}%
\let\ProcessRunnHead=\epjrunnhead
\newpage
%
\maketitle
\section{The muon anomalous magnetic moment}
\label{sec:1}

In this section we give a short overview of the present status of the theory
behind the muon anomalous magnetic moment and a few new results
on the pion loop contribution to the light-by-light part.

Experiment and theory use the anomaly
$a_\mu \equiv (g_\mu-2)/2$.
BNL E821 \cite{exp} dominates the world
average \cite{PDG10} given in Tab.~\ref{tab:g-2}.
The standard model prediction is a bit off. The prediction and
its main parts are listed in Table~\ref{tab:g-2}.
For definiteness we quote numbers and errors of \cite{PDG10},
but there is agreement on all numbers except on
the hadronic light-by-light (HLBL) part.
The difference is given in the last line
of Tab.~\ref{tab:g-2} with errors added quadratically.
The experiment will move to Fermilab with an expected inprovement
of a factor of four. Theory thus needs to improve.
The discrepancy
has created a lot of excitement since many BSM models \emph{can} predict
a value in this range but often a lot more or a lot less. The value of $a_\mu$
provides a major constraint on many BSM models.
Reviews of all aspects are \cite{MRR,JN}.

\subsection{QED, Electroweak and Hadronic Vacuum Polarization}
\label{QED}

The QED contribution is well known. The first term is due to Schwinger.
The first three terms are known analytically.
the fourth is a full
numerical calculation and the fifth is an estimate. With $\bar \alpha = \alpha/\pi$,
\begin{equation}
\label{aQED}
a^\mathrm{QED}_\mu =
0.5\,\bar\alpha
+ 0.765857410(27)\,\bar\alpha^2
+ 24.05050964(43)\,\bar\alpha^2
+ 130.8055(80)\,\bar\alpha^2
+ 663(20)\,\bar\alpha^2
+\cdots
\end{equation}
Kinoshita and collaborators played a major role in evaluating all
contributions numerically. The QED value in Tab.~\ref{tab:g-2} 
uses $\alpha$ from the electron magnetic moment. 
The third order contribution is dominated by the 
unexpectedly large leptonic light-by-light (LLBL)
contribution \cite{AKBD}. The Schwinger diagram
is shown in Fig.~\ref{fig:QED}a and the LLBL
diagram in Fig.~\ref{fig:QED}b with its part of the
$24.05$ QED third order in (\ref{aQED}).

\begin{table}
\begin{minipage}{0.40\textwidth}
\caption{\label{tab:g-2} Overview of results.}
\centerline{
\begin{tabular}{lrc}
\hline\noalign{\smallskip}
$10^{10}a_\mu$
    & value & error \\
\noalign{\smallskip}\hline\noalign{\smallskip}
exp &      11 659 208.9 & 6.3\\
\hline
theory & 11 659 180.2 & 4.9\\
\hline
QED &  11 658 471.8 & 0.0 \\
EW & 15.4 & 0.2\\
LO Had & 692.3 & 4.2 \\
HO HVP &  $-$9.8 & 0.1 \\
HLBL & 10.5 &  2.6 \\
\hline
difference & 28.7 & 8.1\\
\noalign{\smallskip}\hline
\end{tabular}
}
\end{minipage}
\begin{minipage}{0.59\textwidth}
\caption{\label{tab:HLBL} The different parts of the HLBL contribution.}
\begin{tabular}{ccc}
\hline
 & BPP \cite{BPP} & PdRV \cite{PdRV}\\
\hline
pseudo-scalar & $(8.5\pm1.3)\cdot 10^{-10}$ & $(11.4\pm1.3)\cdot 10^{-10}$\\ 
axial-vector & $(0.25\pm0.1)\cdot 10^{-10}$ & $(1.5\pm1.0)\cdot 10^{-10}$\\
quark-loop &  $(2.1\pm0.3)\cdot 10^{-10}$ &  ---\\
scalar &$ (-0.68\pm0.2)\cdot 10^{-10}$ & $(-0.7\pm0.7)\cdot 10^{-10}$\\
$\pi K$-loop & $(-1.9\pm1.3)\cdot 10^{-10}$ & $(-1.9\pm1.9)\cdot 10^{-10}$\\
\hline
errors & linearly & quadratically\\
sum &  $(8.3\pm3.2)\cdot 10^{-10}$ & $(10.5\pm2.6)\cdot 10^{-10}$\\
\hline
\end{tabular}
\end{minipage}
\end{table}

\begin{figure}
\begin{minipage}{0.43\textwidth}
\centerline{
\setlength{\unitlength}{1.2pt}
\begin{picture}(50,65)(0,-15)
\SetScale{1.2}
\SetWidth{0.75}
\Photon(25,50)(25,30){3}{3.5}
\ArrowLine(0,0)(12.5,15)
\ArrowLine(12.5,15)(25,30)
\ArrowLine(25,30)(37.5,15)
\ArrowLine(37.5,15)(50,0)
\Photon(12.5,15)(37.5,15){-2}{4.5}
\Text(25,-10)[]{(a)}
\end{picture}
\hspace*{0.4cm}
\setlength{\unitlength}{1.2pt}
\begin{picture}(60,65)(0,-15)
\SetScale{1.2}
\SetWidth{0.75}
\Photon(30,50)(30,40){3}{2}
\ArrowLine(0,0)(10,0)
\ArrowLine(10,0)(30,0)
\ArrowLine(30,0)(50,0)
\ArrowLine(50,0)(60,0)
\Photon(10,0)(10,15){-3}{3}
\Photon(30,0)(30,15){3}{3}
\Photon(50,0)(50,15){3}{3}
\DashArrowLine(10,15)(30,15){2}
\DashArrowLine(30,15)(50,15){2}
\DashArrowLine(50,15)(30,40){2}
\DashArrowLine(30,40)(10,15){2}
\Text(50,50)[l]{$e=20.95$}
\Text(50,40)[l]{$\mu=0.37$}
\Text(50,30)[l]{$\tau=0.002$}
\Text(30,-10)[]{(b)}
\end{picture}
}
\caption{\label{fig:QED} Examples of QED contributions. (a) The contribution calculated by Schwinger. (b) The leptonic light-by-light contribution.}
\end{minipage}
\hspace{1cm}
\begin{minipage}{0.48\textwidth}
\centerline{
\setlength{\unitlength}{1.2pt}
\begin{picture}(50,65)(0,-15)
\SetScale{1.2}
\SetWidth{0.75}
\Photon(25,50)(25,30){3}{3.5}
\ArrowLine(0,0)(12.5,15)
\DashLine(12.5,15)(25,30){2}\Text(20,23)[rb]{$W$}
\DashLine(25,30)(37.5,15){2}\Text(35,23)[lb]{$W$}
\ArrowLine(37.5,15)(50,0)
\ArrowLine(12.5,15)(37.5,15)
\Text(25,11)[t]{$\nu$}
\Text(25,-10)[]{(a)}
\end{picture}
\hspace*{0.4cm}
\setlength{\unitlength}{1.2pt}
\begin{picture}(60,65)(0,-15)
\SetScale{1.2}
\SetWidth{0.75}
\Photon(30,50)(30,40){3}{2}
\ArrowLine(0,0)(15,0)
\ArrowLine(15,0)(45,0)
\ArrowLine(45,0)(60,0)
\Photon(15,0)(15,15){-3}{3}
\DashLine(45,0)(45,15){2}
\Text(48,7)[l]{$Z$}
\DashArrowLine(15,15)(45,15){2}
\DashArrowLine(45,15)(30,40){2}
\DashArrowLine(30,40)(15,15){2}
\Text(47,45)[l]{$e,\mu,\tau,$}
\Text(47,35)[l]{$u,d,s,$}
\Text(47,25)[l]{$c,t,b$}
\Text(30,-10)[]{(b)}
\end{picture}
}
\caption{\label{fig:EW} The electroweak contributions. (a) A typical 1-loop diagram. (b) An example of a triangle anomaly diagram appearing at 2-loop order.}
\end{minipage}
\end{figure}

A typical one-loop electroweak diagram is Fig.~\ref{fig:EW}a.
Two-loop corrections are large due to
large, partly hadronic, logarithms in diagrams like Fig.~\ref{fig:EW}b,
(triangle) anomaly in (muon) anomaly \cite{CMV,KPPR}.
\begin{equation}
10^{10}a^\mathrm{EW}_\mu = 19.48[1\mathrm{-loop}]
 -4.07(0.10)(0.18)[2\mathrm{-loop}]
 = {15.4}(0.1)(0.2)(\mathrm{triangle})(\mathrm{Higgs~mass})\,.
\end{equation}

The remaining relevant contributions in the standard model are all hadronic.
The largest is the hadronic vacuum polarization (HVP). The bare
quark-loop has large gluonic corrections and needs to be done to all orders
in $\alpha_S$ as depicted in Fig.~\ref{fig:HVP}. This contribution can be related to experiment via
\begin{equation}
\label{aHVP}
a_\mu^\mathrm{LO had} = \frac{1}{3}\left(\frac{\alpha}{\pi}\right)^2
\int_{m_\pi^2}^\infty ds \frac{K(s)}{s}R^{(0)}(s)\,,
\qquad
R^{(0)}(s) \equiv \left.
\frac{\sigma(e^+e^-\to hadrons)}{\sigma(e^+e^-\to \mu^+\mu^-)}\right|_\mathrm{bare}\,.
\end{equation}
The precise definition of bare led to some confusion
between theory and experiment and there were experimental discrepancies. 
A representative value is given in Tab.~\ref{tab:g-2}, see \cite{PDG10,MRR,JN} for references and
discussion.
At higher orders in $\alpha$ two types of hadronic contributions are relevant.
Those with two insertions of the HVP, as
in Fig.~\ref{fig:HOVP}, see~\cite{PDG10,MRR,JN},
can be evaluated from $R^{(0)}(s)$ and
the HLBL contribution is discussed in Sect.~\ref{HLBL}. 
Values are again given in Tab.~\ref{tab:g-2}.

\begin{figure}
\begin{minipage}{0.65\textwidth}
\centerline{
\setlength{\unitlength}{1.2pt}
\begin{picture}(60,55)
\SetScale{1.2}
\SetWidth{0.75}
\Photon(30,55)(30,40){3}{3}
\ArrowLine(0,0)(10,15)
\ArrowLine(10,15)(30,40)
\ArrowLine(30,40)(50,15)
\ArrowLine(50,15)(60,0)
\Photon(10,15)(20,15){-2}{2.5}
\Photon(40,15)(50,15){-2}{2.5}
\SetColor{Red}
\ArrowArc(30,15)(10,180,0)
\ArrowArcn(30,15)(10,180,0)
\end{picture}
\setlength{\unitlength}{1.2pt}
\begin{picture}(60,55)
\SetScale{1.2}
\SetWidth{0.75}
\Photon(30,55)(30,40){3}{3}
\ArrowLine(0,0)(10,15)
\ArrowLine(10,15)(30,40)
\ArrowLine(30,40)(50,15)
\ArrowLine(50,15)(60,0)
\Photon(10,15)(20,15){-2}{2.5}
\Photon(40,15)(50,15){-2}{2.5}
\SetColor{Red}
\ArrowArc(30,15)(10,180,0)
\ArrowArcn(30,15)(10,180,0)
\Gluon(30,25)(30,5){3}{3}
\end{picture}
\raisebox{1cm}{$\Longrightarrow$}
\setlength{\unitlength}{1.2pt}
\begin{picture}(60,55)
\SetScale{1.2}
\SetWidth{0.75}
\Photon(30,55)(30,40){3}{3}
\ArrowLine(0,0)(10,15)
\ArrowLine(10,15)(30,40)
\ArrowLine(30,40)(50,15)
\ArrowLine(50,15)(60,0)
\Photon(10,15)(20,15){-2}{2.5}
\Photon(40,15)(50,15){-2}{2.5}
\GCirc(30,15){10}{0.4}
\end{picture}
}
\caption{\label{fig:HVP} The lowest-order hadronic vacuum-polarization (HVP)
contribution to $a_\mu$. We need to sum all higher order QCD corrections.}
\end{minipage}
\hspace{0.5cm}
\begin{minipage}{0.3\textwidth}
\centerline{
\setlength{\unitlength}{1.2pt}
\begin{picture}(60,55)(0,0)
\SetScale{1.2}
\SetWidth{0.75}
\Photon(30,55)(30,40){3}{3}
\ArrowLine(0,0)(11.25,15)
\ArrowLine(11.25,15)(30,40)
\ArrowLine(30,40)(48.75,15)
\ArrowLine(48.75,15)(60,0)
\Photon(11.25,15)(17,15){-2}{2.5}
\Photon(43,15)(48.75,15){-2}{2.5}
\Photon(27,15)(33,15){-2}{2.5}
\GCirc(22,15){5}{0.4}
\GCirc(38,15){5}{0.4}
\end{picture}
}
\caption{\label{fig:HOVP} A diagram with two insertions of the HVP.}
\end{minipage}
\end{figure}

\subsection{Hadronic Light-by-Light}
\label{HLBL}

\begin{figure}
\begin{minipage}{0.34\textwidth}
\centerline{\includegraphics[width=0.7\textwidth]{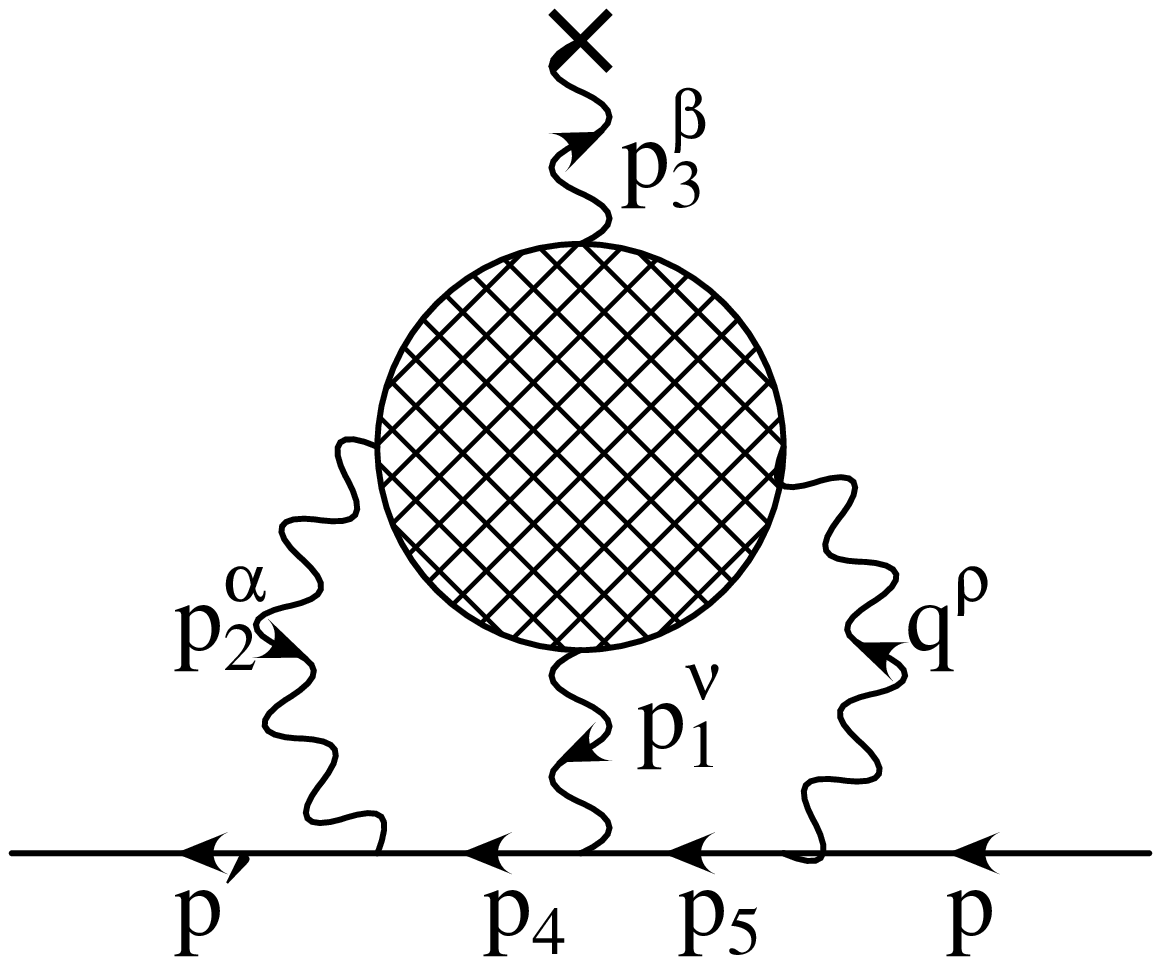}}
\caption{\label{fig:HLBL} The hadronic light-by-light (HLBL) contribution to $a_\mu$.}
\end{minipage}
\hspace{0.2cm}
\begin{minipage}{0.32\textwidth}
\centerline{
\setlength{\unitlength}{1.2pt}
\begin{picture}(100,50)
\SetScale{1.2}
\SetWidth{0.75}
\Photon(0,50)(25,25){3}{5}
\Photon(0,0)(25,25){3}{5}
\Photon(100,50)(75,25){3}{5}
\Photon(100,0)(75,25){3}{5}
\Line(25,27)(75,27)
\Line(25,23)(75,23)
\GCirc(25,25){10}{0.4}
\GCirc(75,25){10}{0.4}
\Text(50,30)[b]{``$\pi^0$''}
\Text(50,10)[b]{$q$}
\Text(10,45)[lb]{$k_1$}
\Text(10,05)[lt]{$k_2$}
\end{picture}
}
\caption{\label{fig:pi0} The pseudo-scalar exchange diagram.
The blobs denote the form-factor $F(q^2,k_1^2,k_2^2)$.}
\end{minipage}
\hspace{0.2cm}
\begin{minipage}{0.28\textwidth}
\setlength{\unitlength}{1.2pt}
\begin{picture}(90,40)
\SetScale{1.2}
\ArrowLine(0,3)(10,3)
\ArrowLine(10,3)(30,3)
\ArrowLine(30,3)(40,3)
\Photon(12,3)(12,40){2}{6}
\Photon(28,3)(28,40){2}{6}
\ArrowLine(50,3)(60,3)
\ArrowLine(60,3)(80,3)
\ArrowLine(80,3)(90,3)
\Photon(62,3)(78,40){2}{8}
\Photon(78,3)(62,40){2}{8}
\end{picture}

\caption{\label{fig:SD} The tree level diagrams that contract to the axial vector current in the limit $P_1^2\approx P_2^2\gg Q^2$.}
\end{minipage}
\end{figure}

The HLBL contribution is depicted in Fig.~\ref{fig:HLBL}. The muon and photon
lines are the well known part. The blob needs to be filled with hadrons
and QCD. The trouble is that low- and high-energy are very mixed
and a double counting of different hadron/quark contributions needs to be
avoided. A workshop at INT, Seattle \cite{INT} provides a
good overview of the situation. A start on separating the different
parts is by studying at which orders
in the large number of colours, $N_c$, and chiral, $p$, expansion, contributions
start \cite{EdR}. The pion loop is $1,p^4$, pion-exchange is $N_c,p^6$
and all others start at $N_c,p^8$. This separation was used to do a full
calculation by two independent groups,
\cite{BPP} and
\cite{HKS}.
The latter used purely hadronic exchanges and added a quark-loop with a
VMD suppression 
as well as the pion loop in hidden local symmetry (HLS) model. 
They studied the dependence on the vector meson mass
to determine the important energy regions. The former used the extended
Nambu-Jona-Lasinio model of \cite{BBR,BENJL} as a basis while repairing
its worst shortcomings. The advantage is that this model has quarks
and automatically generates pseudo-scalars, vectors and axial-vectors
with a reasonable description. 
The scalars are unphysical but describe to some extent
$\pi\pi$ scattering effects.
In addition to the large $N_c$ ENJL contributions
they added the pion-loop with phenomenological VMD in all photon legs
and the short distance quark-loop. They studied the cut-off dependence of the
various contributions. A sign mistake in the pion-exchange by both
groups was discovered by \cite{KN}. Note that since \cite{BPP,HKS}
no new \emph{full} calculation was done.

The HLBL contribution in all detail is given by,
momenta and indices as in Fig.~\ref{fig:HLBL},
\begin{equation}
\label{aHLBLfull}
{a_\mu^{\rm HLBL}} =
{-e^6\over 48 m_\mu} \mathrm{tr}
\!\int {{\rm d}^4 p_1{\rm d}^4p_2 \over (2\pi )^8}
\, {\gamma_\alpha (/\hskip-1.2ex p_4 +m_\mu )\gamma_\nu
(/\hskip-1.2ex p_5 +m_\mu ) \gamma_\rho \over q^2\, p_1^2 \, p_2^2 (p_4^2-m_\mu^2) \,
(p_5^2 - m_\mu^2)}
\left[ {\delta \Pi^{\rho\nu\alpha\beta} (p_1,p_2,p_3)  \over
\delta p_{3\lambda}} \right]
(/\hskip-1.2ex p +m_\mu ) [\gamma_\lambda,\gamma_\beta ]
(/\hskip-1.2ex p +m_\mu ).
\end{equation}
The main object is the four point function
of four electromagnetic currents $V_i^\mu(x) \equiv \sum_i Q_i \, \left[
\bar q_i(x) \gamma^\mu q_i (x) \right]$,
\begin{equation}
\Pi^{\rho\nu\alpha\beta} (p_1,p_2,p_3) \equiv
 i^3 \int {\rm d}^4 x \int {\rm d}^4 y
\int {\rm d}^4 z \,
e^{i(p_1 \cdot x+p_2 \cdot y + p_3 \cdot z)} \,
\langle 0 | T \left( V_a^\rho(0) V_b^\nu(x) V_c^\alpha(y)
V_d^\beta(z) \right) | 0 \rangle
\end{equation}
and we used $
\Pi^{\rho\nu\alpha\lambda}(p_1,p_2,p_3) =
-p_{3\beta} \left(\delta \Pi^{\rho\nu\alpha\beta}(p_1,p_2,p_3)/ \delta  p_{3\lambda}
\right)$
which allows to calculate directly at $p_3=0$ and makes the integrals more
convergent \cite{AKBD}.

The general $\Pi^{\rho\nu\alpha\lambda}$ contains 138 different Lorentz structures
of which 32 contribute to $a_\mu$ \cite{BPP}. Using the gauge invariance
relations $q_\rho \Pi^{\rho\nu\alpha\beta}=
  p_{1\nu} \Pi^{\rho\nu\alpha\beta}=
 p_{2\alpha} \Pi^{\rho\nu\alpha\beta}=
 p_{3\beta} \Pi^{\rho\nu\alpha\beta} = 0$ the 132 can be reduced to 43 structures
\cite{BP,BZ}
that after $p_3\to0$ depend on $p_1^2,p_2^2,q^2$. There are 8 integrals in
(\ref{aHLBLfull}) and most evaluations have rotated the integrations to
Euclidean space. Artefacts in models are smeared out there and the separation
of long and short distances becomes easier.
Three of the integrals are trivial and a new development is that of the
remaining five two can be done using the Gegenbauer polynomial techniques
\cite{JN,KN,BZ}. So in the end integrals over $P_1^2 = -p_1^2,P_2^2 = -p_2^2,
Q^2 = -q^2$ remain. 

To visually see the contribution of various quantities different scales
we introduce \cite{BP}
\begin{equation}
a_\mu^\mathrm{X} = \int dl_{P_1} dl_{P_2} { a_\mu^\mathrm{XLL}}
 = \int dl_{P_1} dl_{P_2} dl_Q { a_\mu^\mathrm{XLLQ}},
\quad\mathrm{with}\quad
 l_P = \ln\left(P/\mathrm GeV\right)\,.
\end{equation}
The contributions of type $X$ at a given scale  $P_1, P_2,Q$
are directly proportional to the volume under the surface when 
$a_\mu^\mathrm{XLL}$ and $a_\mu^\mathrm{XLLQ}$ are plotted 
versus the energies on a logarithmic scale. 

The main contribution is pseudo-scalar exchange, $\pi^0$ 
(and $\eta,\eta^\prime$),
depicted in Fig.~\ref{fig:pi0}. Here one has to model the
form-factor $F(q^2,k_1^2,k_2^2)$ including the dependence on how
off-shell the pion is \cite{JN,BPP,HKS}. Treating it is pointlike gives
a logarithmic divergence \cite{HKS} which can be evaluated using
chiral perturbation theory (ChPT)
\cite{KR,WR}. In \cite{BPP} $\pi^0$-exchange was found to be essentially
saturated at a scale of 1~GeV and 
$10^{10} a_\mu^{\pi^0} = 5.9$. Including $\eta,\eta^\prime$ exchange leads to the value listed under
pseudo-scalar and BPP in Tab.~\ref{tab:HLBL}.
All models except \cite{MV}
give basically a compatible value for $10^{10} a_\mu^{\pi^0}$ with
$6.27$ for the nonlocal quark model \cite{NQM},
 $5.75$ for a Dyson-Schwinger equations based approach
\cite{DSE},
   $(5.8-6.3)$ for a hadronic modeling with two vectors and some short-distance constraints \cite{KN},
$6.54$ for a form-factor inspired by AdS/QCD \cite{ADSQCD},
$(6.5-7.1)$ for the chiral quark model \cite{GR},
$7.2\pm1.2$ with an extra constraint on the form-factor \cite{Nyffeler} and
$7.5$ from a direct AdS/QCD calculation including pion excitations \cite{HK}. 

A new development was the short-distance constraint on
the region $P_1^2\approx P_2^2\gg Q^2$ by \cite{MV}. Here one uses the operator
product expansion of two vector currents to relate $\Pi^{\rho\nu\alpha\beta}$
to a matrix-element of an axial current. \cite{MV} implemented this
constraint by setting one of the form-factors in Fig.~\ref{fig:pi0} to 1, i.e.
pointlike. They obtained a value of $10^{10}a_\mu^{\pi^0} = 7.7$ which
with including $\eta,\eta^\prime$ leads to the number in
Tab.~\ref{tab:HLBL} quoted under pseudo-scalar and PdRV.
The OPE expansion comes from the diagrams of Fig.~\ref{fig:SD}
so one expects that approaches involving the (short-distance)
quark-loop do include this \cite{PdRV,NQM}. The distribution over energy scales
for the two cases is shown in Figs. 5 and 7 of \cite{BP}.

Axial-vector exchange is treated in the same way. \cite{MV} found an
enhancement over \cite{BPP} from both their short-distance constraint and
the mixing of the two axial-vector nonets giving the difference between
the two numbers for that contribution in Tab.~\ref{tab:HLBL}. The ENJL model
of \cite{BPP} gave an estimate for scalar exchange as well.

\begin{figure}
\begin{minipage}{0.49\textwidth}
\centerline{
\includegraphics[width=0.99\textwidth]{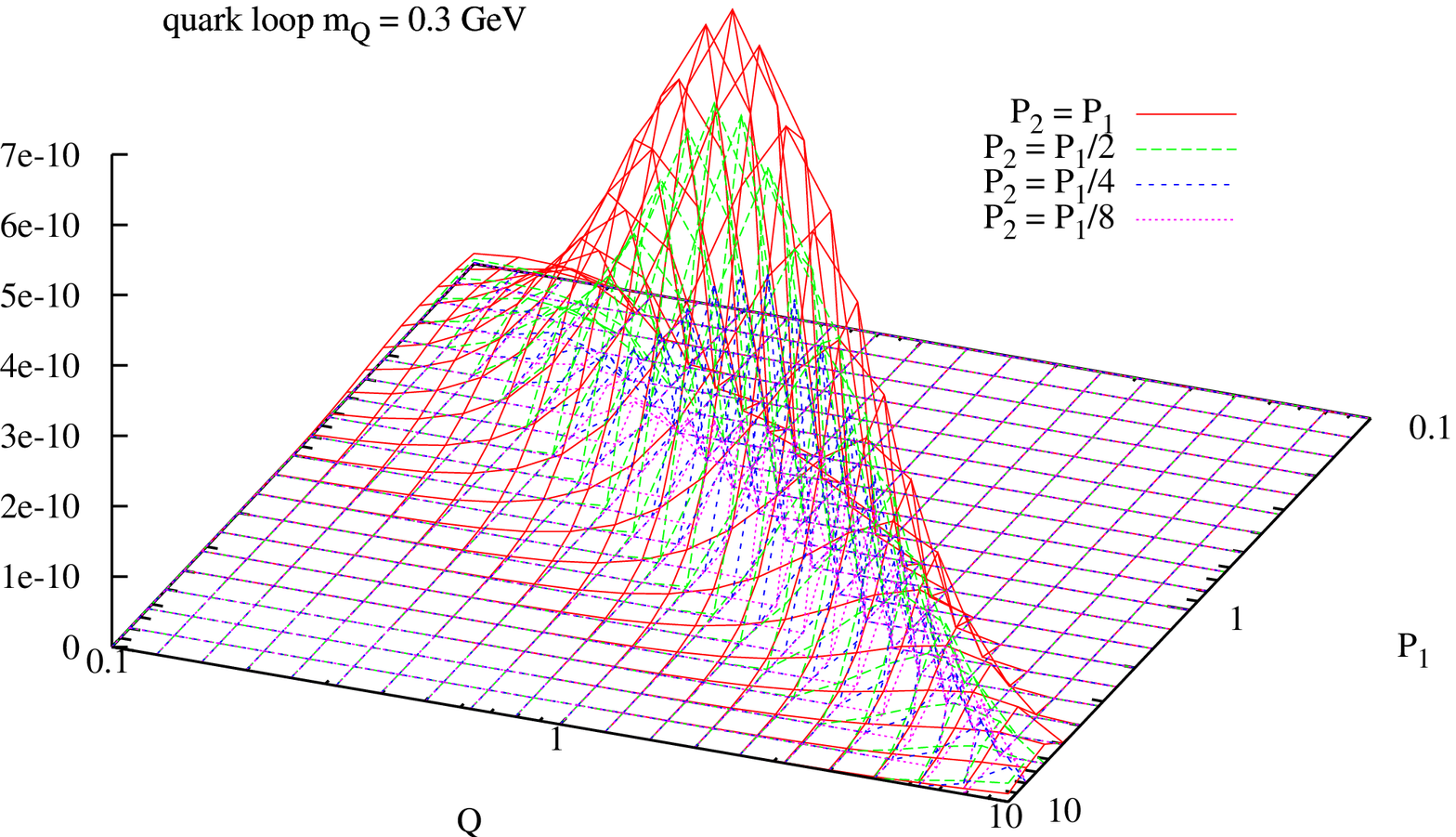}}
\caption{\label{fig:QL} The momentum distribution of the pure quark-loop
contribution for various ratios of $P_2/P_1$.}
\end{minipage}
\begin{minipage}{0.49\textwidth}
\centerline{
\includegraphics[width=0.99\textwidth]{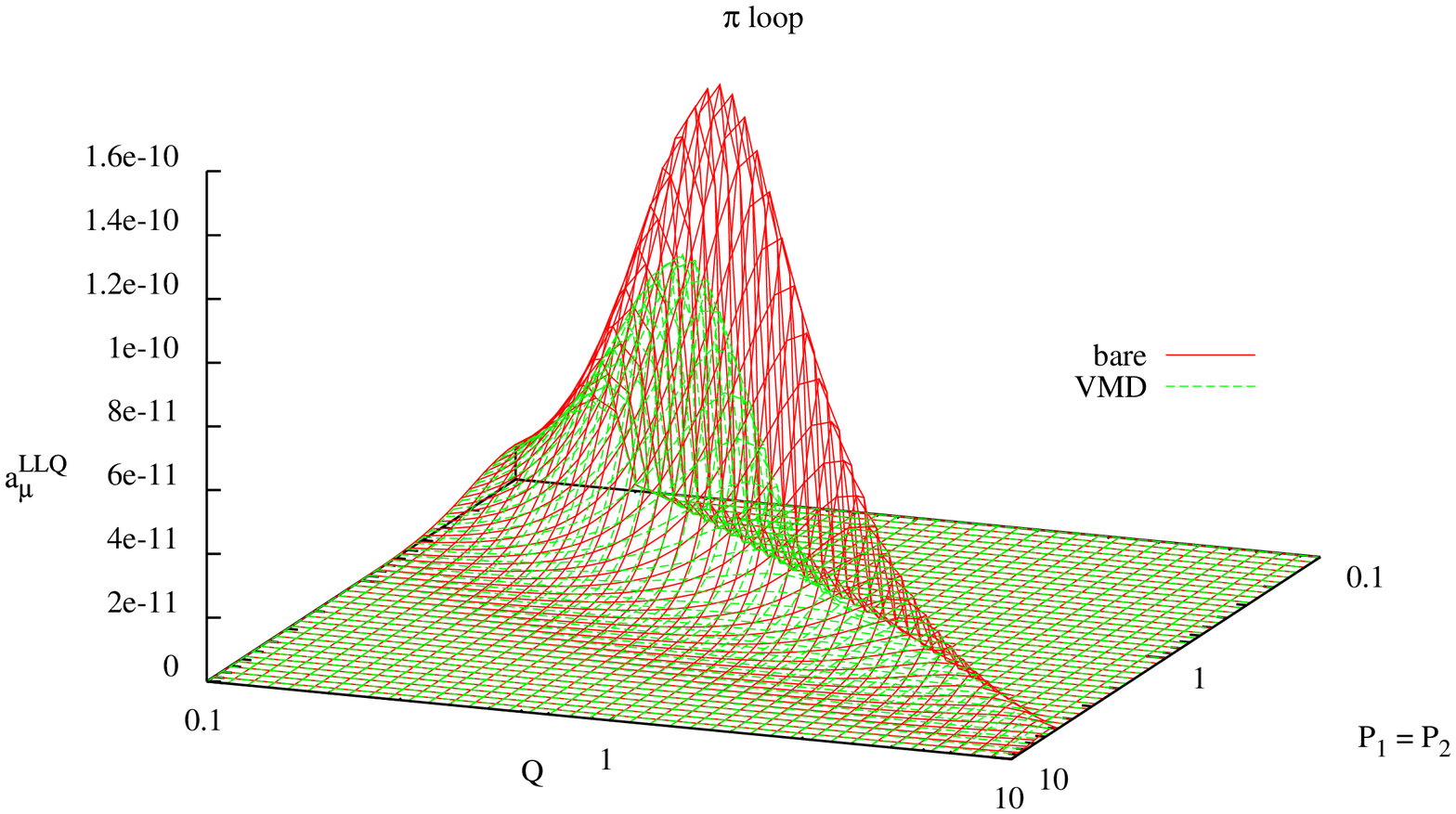}}
\caption{\label{fig:piloop1} The momentum distribution of the bare 
and the pure VMD pion-loop
contribution for $P_2=P_1$.}
\end{minipage}
\end{figure}

The pure quark-loop contribution is known analytically \cite{LR} but estimates
are already much older than that. Using 300~MeV for the quark-mass
one obtains $10^{10}a^{QL}_\mu \approx 5$, with 240~MeV \cite{GR} obtains
about $8$. The contribution over energies is about half below one GeV
\cite{BPP} as can be seen in Fig.~\ref{fig:QL}.
\cite{BPP} used ENJL at low energies and a pure quark-loop with a heavy
quark-mass as cut-off at high energies and found good stability versus
the cut-off with the result quoted in Tab.~\ref{tab:HLBL}.
A much larger value was found in the DSE approach \cite{DSE} but this
calculation needs confirmation.

\begin{figure}
\begin{minipage}{0.49\textwidth}
\centerline{
\includegraphics[width=0.99\textwidth]{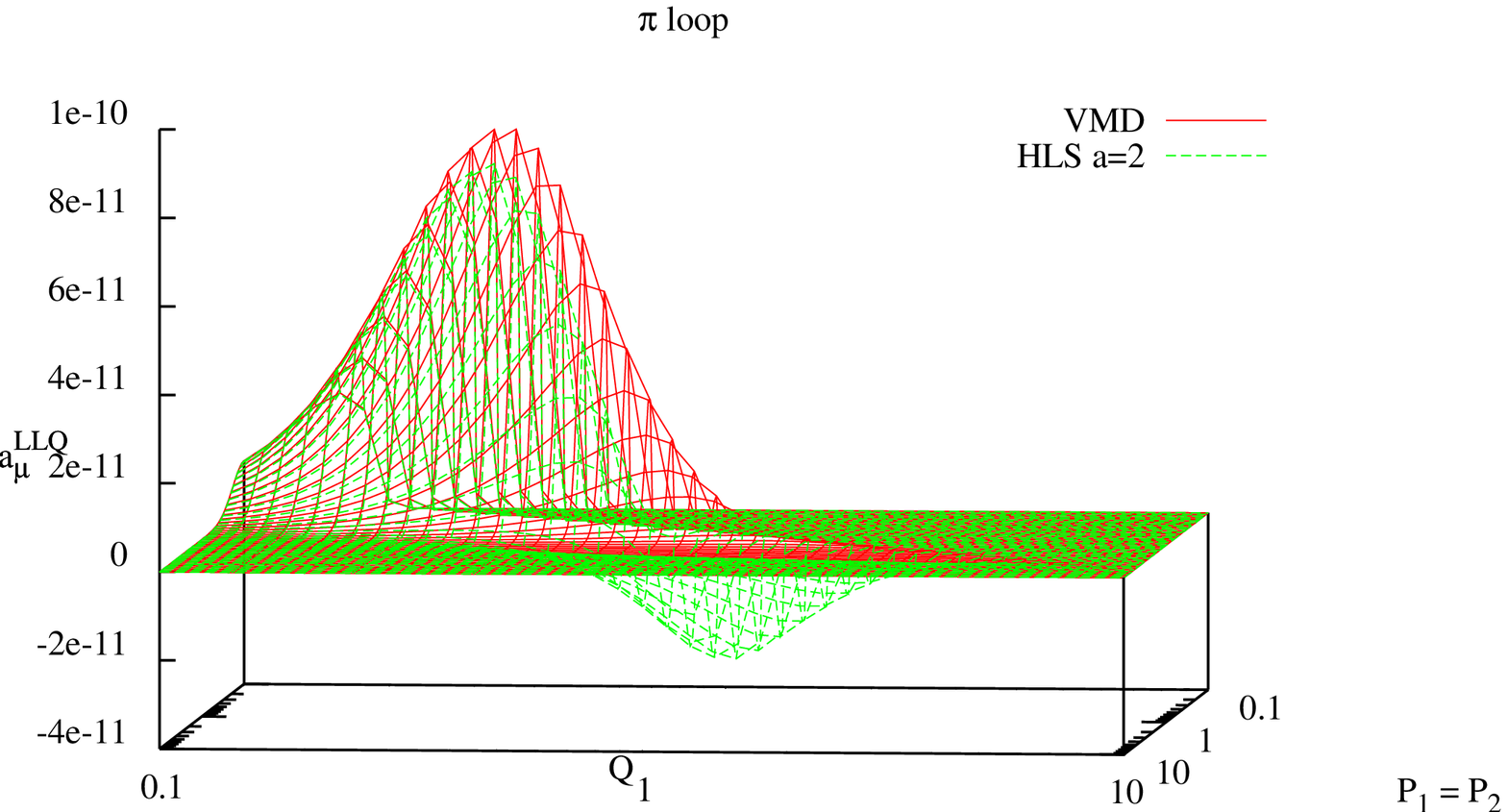}}
\caption{\label{fig:piloop2} The momentum distribution of
the pure VMD and HLS pion-loop
contribution for $P_2=P_1$.}
\end{minipage}
\begin{minipage}{0.49\textwidth}
\centerline{
\includegraphics[width=0.99\textwidth]{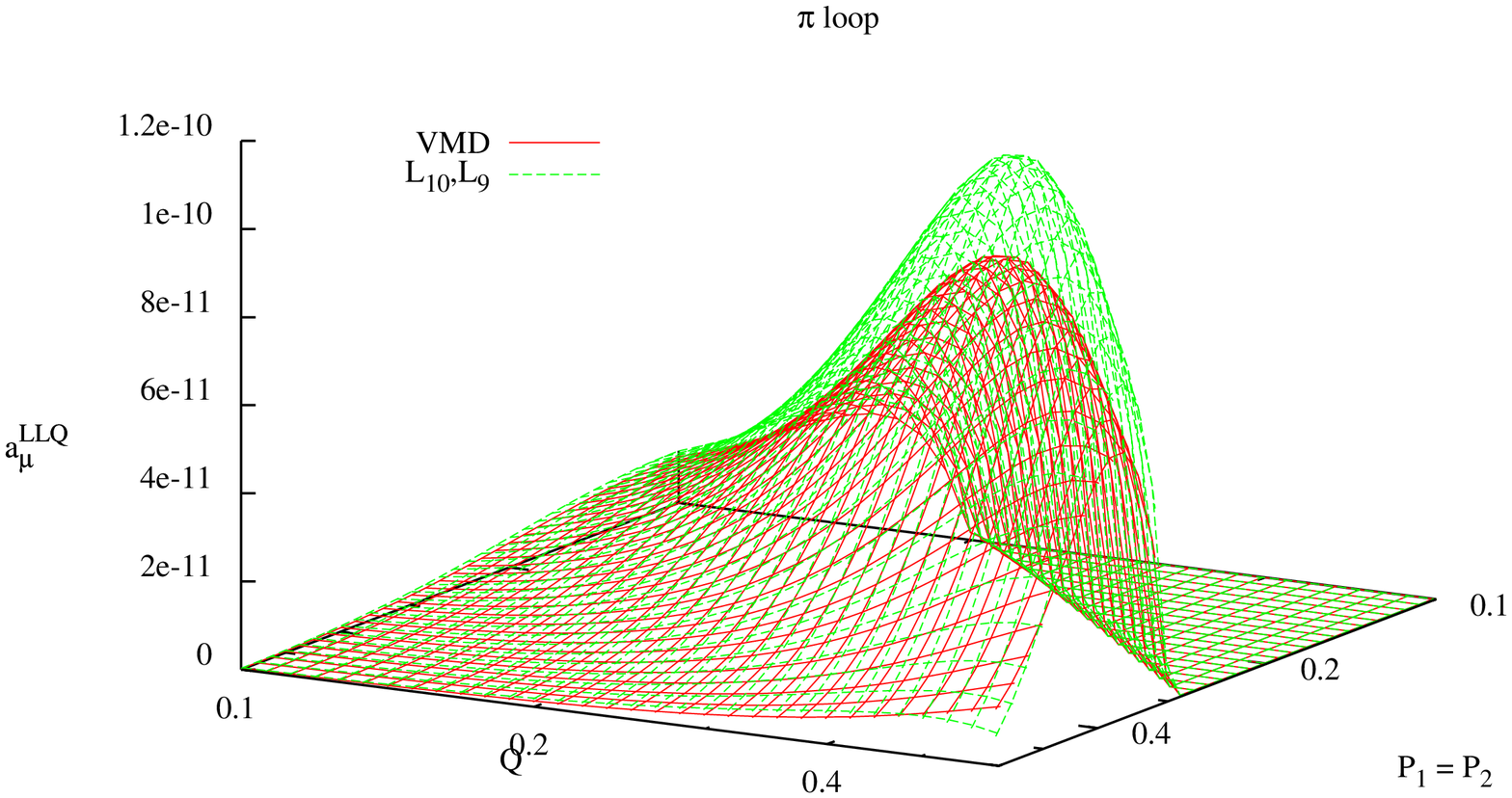}}
\caption{\label{fig:piloopL9L10} The momentum distribution of
the pure VMD pion-loop and the bare pion loop with $L_9,L_{10}$ effect included,
contribution for $P_2=P_1$.}
\end{minipage}
\end{figure}

The last contribution is the one leading in ChPT, the charged pion,
and a small kaon, loop. The bare pion loop gives a large contribution of
$10^{10} a^{\pi\mathrm{-loop}}=-4.1$ but this is expected to be too large.
Several chirally invariant models were used, the HLS approach \cite{HKS}
which gave  $-0.45$ and the VMD inspired approach \cite{BPP} which gave
$-1.9$, an exact VMD approach gives $-1.6$ \cite{BPP,HKS,BZ}.
One can derive from the OPE of two vector currents also a short distance
relation for the $\gamma^*\gamma^*\pi\pi$ process which the HLS and the bare
vertex do not satisfy while the VMD inspired approaches do \cite{BZ}.
The distribution over momenta of the contributions is shown in
Fig.~\ref{fig:piloop1} for the bare and the pure VMD case and in
Fig.~\ref{fig:piloop2} for the HLS and pure VMD.
Notice how the large momentum contributions are cut-off in both the VMD and
HLS case and the source for the large difference between VMD and HLS is the
negative contribution at larger momenta for the HLS indicating that the VMD
result is probably more correct.

\cite{EPR} calculated the pion-loop four-point function
$\Pi^{\rho\nu\alpha\beta}$ at very low momenta in ChPT.
They found that for $P_1,P_2,P_3,Q \ll m_\pi$ there are indications that
the effect due to $L_9$ and $L_{10}$ which in the HLS and VMD models is only
partially taken into account could be important. We show the
pure VMD result with the pion loop result including the effect of 
$L_9$ and $L_{10}$ for scales up to 500~MeV in Fig.~\ref{fig:piloopL9L10}.
We did not take the limit $P_1,P_2,P_3,Q \ll m_\pi$. 
One sees indeed an enhancement of 10\% due $L_9,L_{10}$.
The full contribution to $a_\mu$ at that order in ChPT is divergent.

We conclude like \cite{JN,BPP,HKS,PdRV} that
$a^\mathrm{HLBL}\approx(10\pm3)\cdot10^{-10}$ where the exact central value and error are somewhat subjective.

\section{Renormalization group for EFT}
\label{sec:RGE}

Take an observable $F$ that depends on a single scale
$M$. The dependence on this scale in quantum field theory (QFT)
is typically logarithmic
\begin{equation}
F(M) = F^0_0+F^1_1 L + F^1_0 +\sum_{n=2,\infty}\sum_{m=0,n}F^n_m L^m,
\qquad L=\ln(\mu/M)\,.
\end{equation}
The leading logarithms, the terms $F^m_m L^m$, are easier to calculate
than the full result. This follows from two facts.
The dependence of any observable on the subtraction scale $\mu$
vanishes, $\mu\, dF/d\mu=0$ and ultraviolet divergences
in QFT
are local. In renormalizable QFT
leading logarithms can be described by a running coupling. This can be
proven directly from the renormalization group, but relies on the fact that
in a renormalizable theory counter-terms are of the same form at every order.
It implies that the leading logarithms are calculable by a
one-loop calculation.
The counter-terms for an effective low-energy theory, e.g. ChPT,
differ at every order. 
However, Weinberg \cite{Weinberg0} pointed out
that the two-loop leading logarithms can still be calculated from a one-loop
calculation. The all order generalization was proven 
using beta-functions \cite{BC0} and diagrams \cite{BC1} .
The main underlying observation is that at $n$-loop order all
divergences must cancel. Using dimensional regularization with $d=4-w$
the coefficients of
\begin{equation}
\{1/w^n,\log\mu/w^{n-1},\log^2\mu/w^{n-2},\ldots,\log^{n-1}/w\};~
\{1/w^{n-1},\log\mu/w^{n-2},\ldots,\log^{n-2}\mu/2\};\ldots
\end{equation}
must cancel. 
The first set of conditions allows to prove that all leading logs can
be determined from one-loop diagrams, the second set that the subleading
logs can be had from two-loop diagrams, etc..

The observation \cite{BC1} that the needed Lagrangians at each order
do not need to be minimal, allows them to be computer generated. The
number of diagrams increases fast with order, e.g. mass at six loops
requires 303. The size of each diagram grows even faster.
The leading logarithms for the
mass in the massive $O(N)$ model were calculated to five loops
in \cite{BC1}, for the decay constant and vev to the same order
and for the vector and scalar form-factors as well as meson-meson scattering
to four-loops in \cite{BC2}. 
\cite{BC1,BC2} also discussed the large $N$ limit
for these quantities.
The mass, decay constant and vector
form-factor were pushed to six loops in \cite{BKL}.
\cite{BKL}s main purpose was including the anomaly for the massive $O(3)$
model, i.e. two-flavour ChPT.
The amplitude for $\pi^0\to\gamma^*(k_1)\gamma^*(k_2)$ is
\begin{equation}
 A_{\pi\gamma^*\gamma^*} = \epsilon_{\mu\nu\alpha\beta}\,
\varepsilon_1^{*\mu}(k_1)\varepsilon_2^{*\nu}(k_2)\,k_1^\alpha k_2^\beta \, 
F_{\pi\gamma\gamma}(k_1^2,k_2^2)\quad
F_{\pi\gamma\gamma}(k_1^2,k_2^2) = \frac{e^2}{4\pi^2 F_\pi}\hat F
 F_\gamma(k_1^2) F_\gamma(k_2^2)
 F_{\gamma\gamma}(k_1^2,k_2^2)\,.
\end{equation}
$\hat F$ is there for on-shell photons; $F_\gamma(k^2)$ is the form-factor
for one off-shell photon and $F_{\gamma\gamma}$ is the nonfactorizable part
when both photons are off-shell.
The leading logarithms to six loops \cite{BKL}
are numerically
\begin{equation}
\hat F = 1+0-0.000372+0.000088+0.000036+0.000009+0.0000002+\cdots
\end{equation}
showing extremely good convergence.
The form-factor $F_\gamma(k^2)$ also converges well but here the leading
logarithms are known to be only a small part.
The nonfactorizable part $F_{\gamma\gamma}$
only starts at three-loop order (could have started at two)
and in the chiral limit only starts at four-loops.
The leading logarithms thus predict this part to be fairly small.

Similarly, the leading logarithms for the $\gamma3\pi$ vertex
are small and give a good convergence with
\begin{equation}
F_0^{3\pi LL} = (9.8-0.3+0.04+0.02+0.006+0.001+\cdots) \; \mathrm{GeV}^{-3}\,.
\end{equation}

\section*{Acknowledgements}

Supported in part by the European Community
SP4-Capacities
(HadronPhysics3, Grant Agreement n. 283286)
and the Swedish Research Council grants 621-2011-5080 and 621-2010-3326

\end{document}